# Interaction-Driven Spectrum Reconstruction in Bilayer Graphene

A. S. Mayorov[1], D. C. Elias[1], M. Mucha-Kruczynski[2], R. V. Gorbachev[3], T. Tudorovskiy[4], A. Zhukov[3], S. V. Morozov[5], M. I. Katsnelson[4], V. I. Fal'ko[2], A. K. Geim[3], K. S. Novoselov[1]

[1]*School of Physics & Astronomy, University of Manchester, Manchester M13 9PL, UK*

[2]*Physics Department, Lancaster University, Lancaster LA1 4YB, UK*

[3]*Manchester Centre for Mesoscience & Nanotechnology, University of Manchester, Manchester M13 9PL, UK*

[4]*Radboud University of Nijmegen, Institute for Molecules and Materials, Heyendaalseweg 135, 6525 AJ Nijmegen, The Netherlands*

[5]*Institute for Microelectronics Technology, 142432 Chernogolovka, Russia*

**The nematic phase transition in electronic liquids, driven by Coulomb interactions, represents a new class of strongly correlated electronic ground states. We studied suspended samples of bilayer graphene, annealed so that it achieves very high quasiparticle mobilities (greater than $10^6$ cm$^2$/V·s). Bilayer graphene is a truly two-dimensional material with complex chiral electronic spectra and the high quality of our samples allowed us to observe strong spectrum reconstructions and electron topological transitions that can be attributed to a nematic phase transition and a decrease in rotational symmetry. These results are especially surprising because no interaction effects have been observed so far in bilayer graphene in the absence of an applied magnetic field.**

Although many consequences of the chiral nature of the quasiparticles in mono-[1-4] and bi-layer[5-6] graphene have been demonstrated already[3, 7-10], sample inhomogeneities and broadening of the energy levels (due to a finite lifetime of the charge carriers) prohibited a direct access to the intriguing regime of zero electron and hole concentrations. The latter regime has attracted a large number of possible scenarios of the particular arrangement of the electron ground state. For bilayer graphene, the single electron picture[6] (where parabolic spectrum is replaced by four Dirac cones at low energies) is challenged by several many-body phase transition scenarios: ferromagnetic[11]; ferroelectric[12-13] and nematic[14-15]. The latter resembles the transitions in correlated anisotropic electron fluids [16-19] which has been observed in several systems, including electrons in high Landau levels in two-dimensional electron gas[20-22], several layered complex oxides[23-24], iron pnictides[25] and strontium ruthenate[26].

Recently, the required access to the low-energy physics in bilayer crystals has been enabled by improvements in the quality of samples created by removing the underlying substrate[27-29]. Free-standing graphene devices allow a 100-fold increase in the mobility (above $10^6$ cm$^2$/V·s) [30] after annealing. We studied the low-energy electronic properties of such high-quality, free-standing bilayer graphene devices and demonstrate that e-e interactions effects play a dominant role at low energies, even at low magnetic fields strengths B. Our samples were prepared on Si/SiO$_2$ substrate (300nm of SiO$_2$) by micromechanical cleavage technique[1]. Narrow (2 to 4μm) ribbons were selected, and Cr/Au (5nm/150nm) contacts were deposited by means of electron-beam or laser-writer lithography and electron-beam evaporation to create two-probe devices. Wet etching in buffered hydrofluoric acid has been used to remove about 150nm of SiO$_2$ substrate to form free-standing bilayer graphene devices (Fig. 1C).

As prepared, such samples were p-doped to about several $10^{12}$ cm$^{-2}$ and exhibited mobility traditionally expected for samples on a substrate[31] (~5,000 cm$^2$/V·s). We used current annealing[28, 32] in order to remove such unintentional doping and increase the mobility (typically 0.5-1 mAmps/μm current densities were used). After annealing, our devices exhibited a very sharp resistance peak situated practically exactly at zero gate voltage (within $10^8$ cm$^{-2}$ in terms of residual doping; Fig. 1, A and B). Typical quantum mobilities achieved were within $\mu_q$=0.5-1.5×$10^6$ cm$^2$/V·s (estimated from the onset of quantum oscillations in magnetic field $B_0$ as $\mu_q B_0$=1). Transport mobilities were more difficult to estimate because of the influence of the contacts in the two-probe geometry and nearly ballistic transport in our high-quality samples.

However, a rough estimate made from the slope of the conductivity curve (field effect mobility: $\mu=1/e \cdot d\sigma/dn$, where $\sigma$ is conductivity) gave similar numbers (within 30%).

Our experiments on electronic transport in free-standing bilayer graphene in zero and finite magnetic fields show that at energies below 5 meV the parabolic dispersion (as usually observed at high energies[5]) is replaced, through an electron topological transition, by several pockets with linear spectrum. Also, our magnetic field measurements show the presence of only two Dirac cones in the vicinity of each K point in the Brillouin zone, which is in clear contrast with the single-particle theory[6] that predicts four Dirac cones in the vicinity of each K point at energies below 1 mev. We interpret the results of our experiments as an indication for interaction-driven phase transition[14].

All of the samples described in this Report had practically zero doping (within $10^8 cm^{-2}$), which allows us to estimate the possible energy gap that would arise from asymmetry between the layers. This gap is lower than 0.1meV, which is an order of magnitude smaller than the effects we describe. The absence of the gap is also manifested in the relatively weak temperature dependence of the resistivity peak (conductivity minimum) down to the lowest temperatures measured ($T$=0.25K) (Figs. 1A and 2A). At low temperatures, the conductivity saturates at a finite value of about $20e^2/\pi h$ (see figure S1[33] for details), which is substantially greater than the value expected for ballistic monolayer graphene[33-36] ($4e^2/\pi h$) or bilayer graphene with a parabolic dispersion relation[37] ($8e^2/\pi h$). The measured minimum conductivity is slightly lower than $24e^2/\pi h$ predicted[38] for bilayer spectrum transformed by the electron topological transition at low Fermi energies[6] (Fig. 2B).

The broadening and the amplitude of the conductivity minimum were linear functions of temperature for $T$>20K. Broadening of the conductivity minimum effectively yielded the same information as the amplitude, but without the ambiguity which arises from the contact resistance and the quantum minimum conductivity (see Supplementary Information[33] for the definition of broadening we use here). Such behavior is expected for the parabolic dispersion relation, which guarantees a constant density of states (conductivity is: $\sigma_{min}=e\mu \int D(E)[f(E,T)-f(E,T=0)]dE = 4ln(2)e\mu mT/\pi \hbar^2$, where $D(E)$=const – is the density of states, $f(E,T)$ – the Fermi-Dirac distribution function, $m$ - the effective mass, and we assumed that the mobility is independent of the temperature and the position of the Fermi level). The slope of the temperature dependence yielded $m \approx (0.029 \pm 0.003)m_0$, Fig. 2B (here $m_0$ is the free electron mass). Note, that according to the same arguments, the conductivity for the monolayer graphene and broadening of the conductivity minimum increased much slower and (since $D(E) \propto |E|$ for the linear spectrum) is a parabolic function of $T$ (Fig. 1B and Fig. 2, B and C).

For $T<20K$, the temperature dependence of the width of the conductivity minimum for bilayer was substantially weaker than what is expected for the parabolic dispersion relation (the amplitude of the conductivity minimum yields similar conclusion). A slow $T$-dependence hints that the density of states at low energy is suppressed, although it remains finite. Such behavior has been predicted[6] for the low-energy part of the bilayer spectrum, below the saddle-point $E^*$ in the electron dispersion, where the high-energy parabolic spectrum is replaced by four Dirac mini-cones (see inset Fig. 2B); the value of $E^*$ is determined by the amplitude of the next-nearest neighbor interlayer hopping $\gamma_3$, and the number of pockets by the trigonal symmetry of this parameter.

A change in the band structure topology, which consists in a single-connected isoenergetic line splitting into disconnected parts, is known as the electron topological or Lifshitz transition[39]. In bilayer graphene, such a transition is expected to occur at $E^*\sim1meV$, yielding a temperature dependence presented on Fig. 2A (green curve). Although such theoretical prediction qualitatively catches the general trend (transition from parabolic to linear temperature dependence), it strongly underestimates the value of $E^*$. It requires $E^*$ of the order of 6 meV to fit our experimental data, see Fig. 2A (red curve, see below for the details of the model used).

To study the electron topological transition in more detail and search for evidence of many-body effects, we measured the transport properties of such high-quality bilayer samples in an applied magnetic field. Our measurements are summarized on Fig. 3, where the resistivity is plotted as a function of magnetic field and gate voltage. We observed non-monotonic changes in resistivity at filling factor zero, $\nu=0$, Fig. 3B. The sign of magnetoresistance changed twice: it was positive at low (<0.1T) and high (>0.3T) magnetic fields and negative in the intermediate region. Generally, higher quality samples exhibited these features at lower fields. We interpret the divergence of the resistance at high fields ($B$>0.3T) as the gap opening at $\nu=0$, similar to that observed previously[29, 40]. The negative magnetoresistance at 0.1T<$B$<0.3T has a resonant character – a narrow cusp appears in resistance in a narrow area of concentrations ($\pm2\times10^9$cm$^{-2}$) (Fig. 3C, Fig. S2 and S3[33]), an effect that we cannot explain at present. Data for a device with different mobility is shown in figure S2[33].

We also observed that the cyclotron gaps at filling factor $\nu=\pm 4$ (the filling factor is determined from the carrier concentration $n$ as $\nu=nh/eB$), was much more robust than the gaps at other filling factors and could be observed at magnetic fields as low as 100G (the other filling factors are observed only at $B>0.4$T even for the highest quality samples, which is consistent with the results obtained by other groups[40]). Fig. 4 summarizes our measurements of the cyclotron gaps at various filling factors (obtained from fitting of the temperature dependence of oscillations presented on Fig. 3C with Lifshitz-Kosevich formula). The cyclotron gaps at filling factors $\pm 8$, $\pm 12$, $\pm 16$, $\pm 20$ and $\pm 24$ showed a linear dependence on magnetic field (Fig. 4). We extracted the corresponding cyclotron mass, which appears to be $m=(0.028\pm 0.002)m_0$ – very close to previously reported value for non-suspended samples[41]. At the same time, the gap for $\nu=\pm 4$ rose quickly for $0$T$<B<0.02$T and then stayed practically constant at $E_L\approx 5.5$meV for $0.02$T$<B<0.8$T, before showing a linear increase with a slope approximately 40% greater than that for other filling factors.

The reported observation cannot be explained within the framework of parabolic bands, which predicts roughly equidistant Landau levels. It also contradicts the more elaborative theory[6] (beyond the nearest neighbor hopping approximation) which predicts four Dirac mini-cones below the electron topological transition energy in the vicinity of both K and K' points. Similarly to graphene, each of such Dirac pockets would result in doubly-degenerate Landau level, so the total degeneracy of the zero Landau level is 16, and the largest cyclotron gap should correspond to filling factor $\pm 8$, instead of $\pm 4$ as observed in our experiments.

To explain the observed 8-fold degeneracy of the lowest Landau level, we need to assume that the rotational symmetry of the system is lowered by, *e.g.,* nematic phase transition[14-15], which results in the following Hamiltonian for the low-energy electrons,

$$H = -\frac{1}{2m}\begin{pmatrix} 0 & (\pi^\dagger)^2 \\ \pi^2 & 0 \end{pmatrix} + \xi v_3 \begin{pmatrix} 0 & \pi \\ \pi^\dagger & 0 \end{pmatrix} + \begin{pmatrix} 0 & u \\ u & 0 \end{pmatrix}.$$

and reduces the number of low-energy Dirac mini-cones from four to two near each of the two K-points[14-15], Fig. 4A. In the above Hamiltonian, $\xi$ differentiates between the two valleys, $v_3$ is the velocity associated with the interlayer skew hopping[6] and $u$ is the order parameter formed upon nematic phase transition[14-15]. This transition is driven by Coulomb interaction between electrons (which strength is partly reduced by screening), and it has been established by the renormalization-group study of bilayer parameters and the strength of the interaction between all types of symmetry-breaking fluctuation in the electronic liquid in bilayer graphene[14-15]. We could fit our experimental results for the cyclotron gaps rather accurately (Fig. 4, B and C). The range of parameters that fit our results was very narrow, and the same set of parameters fit our temperature broadening data at zero magnetic field (Fig. 2, A and B).

Lowering of the symmetry can be expected either in the case of the spontaneous symmetry breaking caused by many body effects or, also, in the case of mechanical deformation of our devices: either by uniaxial strain of the whole crystal[42-43] or by inter-layer shear shift[44] caused by their preparation/handling history. However, the latter scenario can be ruled out because the strain required to explain the results observed in our work could be estimated[42-43] to be of the order of 1%, whereas the typical strain found in similar devices (prepared and handled in the same way) was measured[30] was below 0.1%. Moreover, in the case of externally induced strain, the direction of mechanical deformations would be related to the contact geometry, not to the crystallographic orientation, and would vary from sample to sample. Instead, we find the same feature in 4 different devices, which would require rather narrow range of parameters (size and orientation of the principal axes of strain tensor) to be repeated precisely in different samples. We also have considered a possibility to explain our results in terms of spontaneous gap opening in bilayer spectrum[45]. In this case the transport at the neutrality point (with or without magnetic field) would be dominated by a network of one-dimensional channels between the domains of gaps with alternating signs. This would result in a minimum conductivity value which depends strongly on the prehistory, in contrast to values reproducibly observed in our experiments. Also, it is hard to expect a deep in the resistance at $\nu=0$ in such model, as observed in our experiments. This suggests that the observed reconstruction of the spectrum is the result of an intrinsic modification of the electronic system, in particular, it can be caused by the recently predicted 'nematic' phase transition[14-15] in bilayer graphene.

*Supported by the European Research Council, European Commision FP7, Engineering and Physical Research Council (UK), the Royal Society, U.S. Office of Naval Research, U.S. Air Force Office of Scientific Research, and the Körber Foundation.*

Tel.: +44-(0)161-275-41-19 (office)
Fax.: +44-(0)161-275-40-56
Web.: www.kostya.graphene.org
E-mail: kostya@manchester.ac.uk

**Figure Legends**
**Figure 1** Temperature dependence of the resistance in graphene and bilayer graphene. (A) Temperature dependence of two terminal resistance of bilayer graphene at zero magnetic field. (B) Temperature dependence of two probe resistance of monolayer graphene at zero magnetic field. (C) An SEM image of one of our suspended devices.

**Figure 2** Temperature dependence of the width and the amplitude of the conductivity minimum for mono- and bi-layer free-standing graphene samples. (A) Width of the conductivity minimum for bilayer graphene (open circles and open squares – experiments for two different samples, red line – theory for bilayer graphene with low-energy spectrum reconstructed due to nematic phase transition, green line – theory for bilayer with non-reconstructed spectrum). Fitting parameters for the reconstructed spectrum (red line): $m=0.0280m_0$, $v_3=1.41 \times 10^5$ m/s, $u=-6.32$meV; those for non-reconstructed (green line): $m=0.0280m_0$, $v_3=1.41 \times 10^5$ m/s, $u=0$. Inset: amplitude of the conductivity minimum of bilayer graphene (yellow crossed circles – experiment; yellow solid line – a guide for an eye). Note a deviation from the straight line below 10K (marked by arrow). (B) The broadening of the conductivity minimum for bilayer samples (circles, squares, red and green lines – same as in A) and for monolayer graphene (blue line - theory). Insets: left – low-energy electronic spectrum as expected in the single-electron approximation; right – bilayer graphene low-energy electronic spectrum, reconstructed due to nematic phase transition. (C) The broadening of the conductivity minimum for monolayer graphene (blue and green triangles - experimental points for two different samples, blue line – theory, same as in B). Inset: low-energy electronic spectrum for monolayer graphene.

**Figure 3** Free-standing bilayer samples in magnetic field. (A) Resistance of suspended bilayer graphene as a function of gate voltage (recalculated to carrier concentration) for B=0.01T (green); B=0.2T (orange) and *B*=0.26T (dark yellow). (B) Resistance of suspended bilayer graphene at $\nu$=0 as a function of magnetic field. (C) Same as A for a greater span of the carrier concentration. (D) Contour map of resistance of free-standing bilayer graphene device as a function of carrier concentration and magnetic field. The resistance spans from 2.1kΩ (blue) to 6kΩ (red). The three vertical lines corresponds to the positions of the B=0.01T (green), B=0.2T (orange) and *B*=0.26T (dark yellow) on A and C.

**Figure 4** (A) Low-energy electron spectrum for graphene bilayer, reconstructed due to nematic phase transition. (B) Cyclotron gaps for different filling factors. Symbols (experiment) and curves (theory) of the same color correspond to the same filling factor (the fitting parameters we used: *m*=0.0280$m_0$, $v_3$=1.41x10$^5$ m/s, *u*=-6.32meV). (C) The position of the Landau levels as a function of magnetic field, recalculated from B. Symbols (experiment) and curves (theory) of the same color correspond to the same filling factor (same as in B).

1 **Figures**

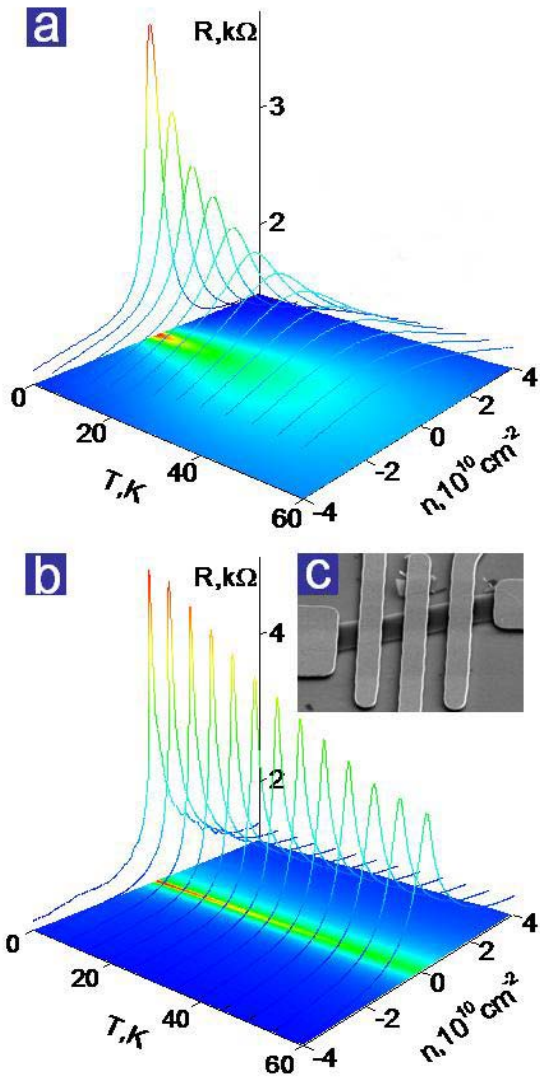

3 Fig. 1

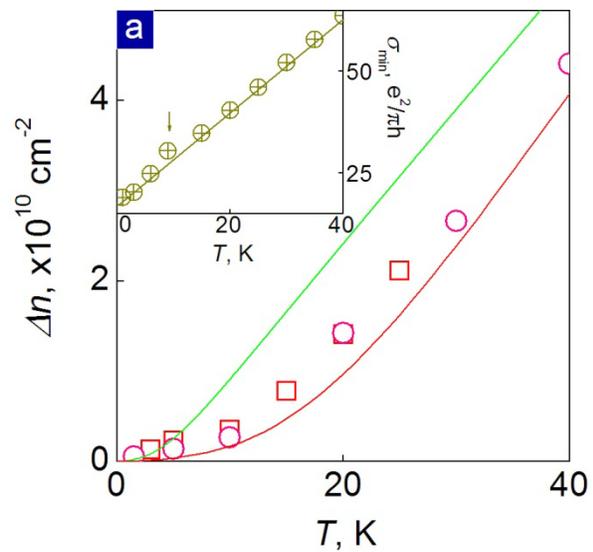
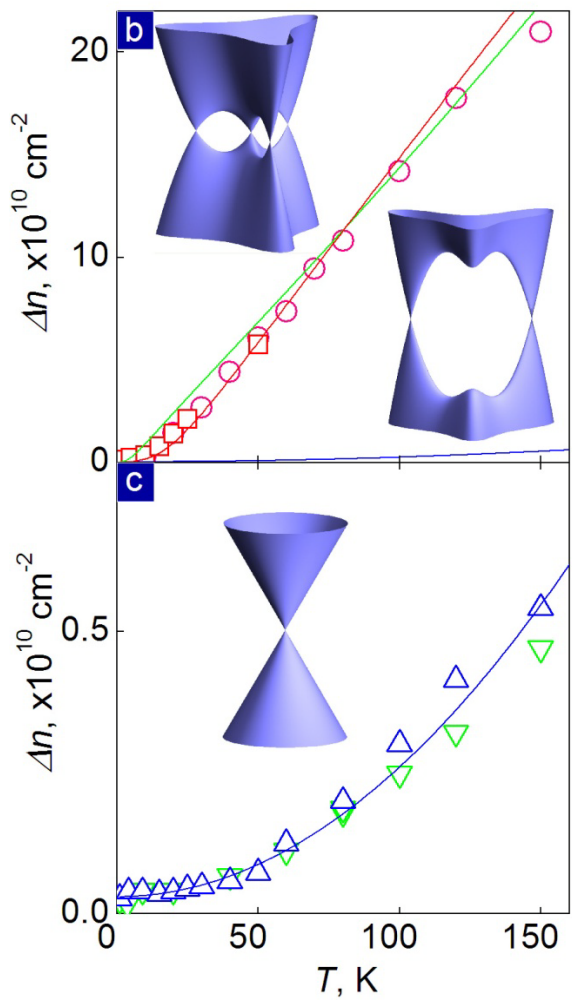

3   Fig. 2

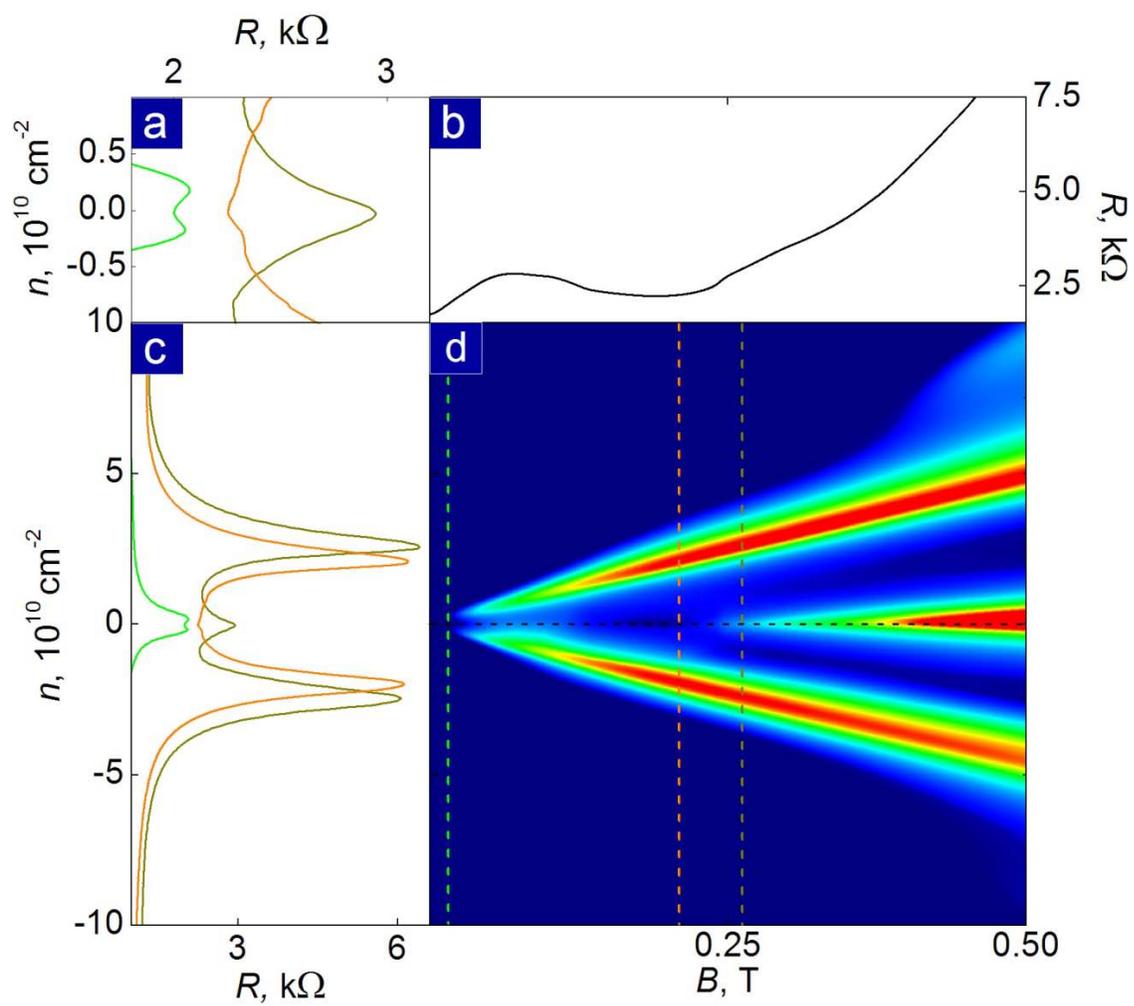

Fig. 3

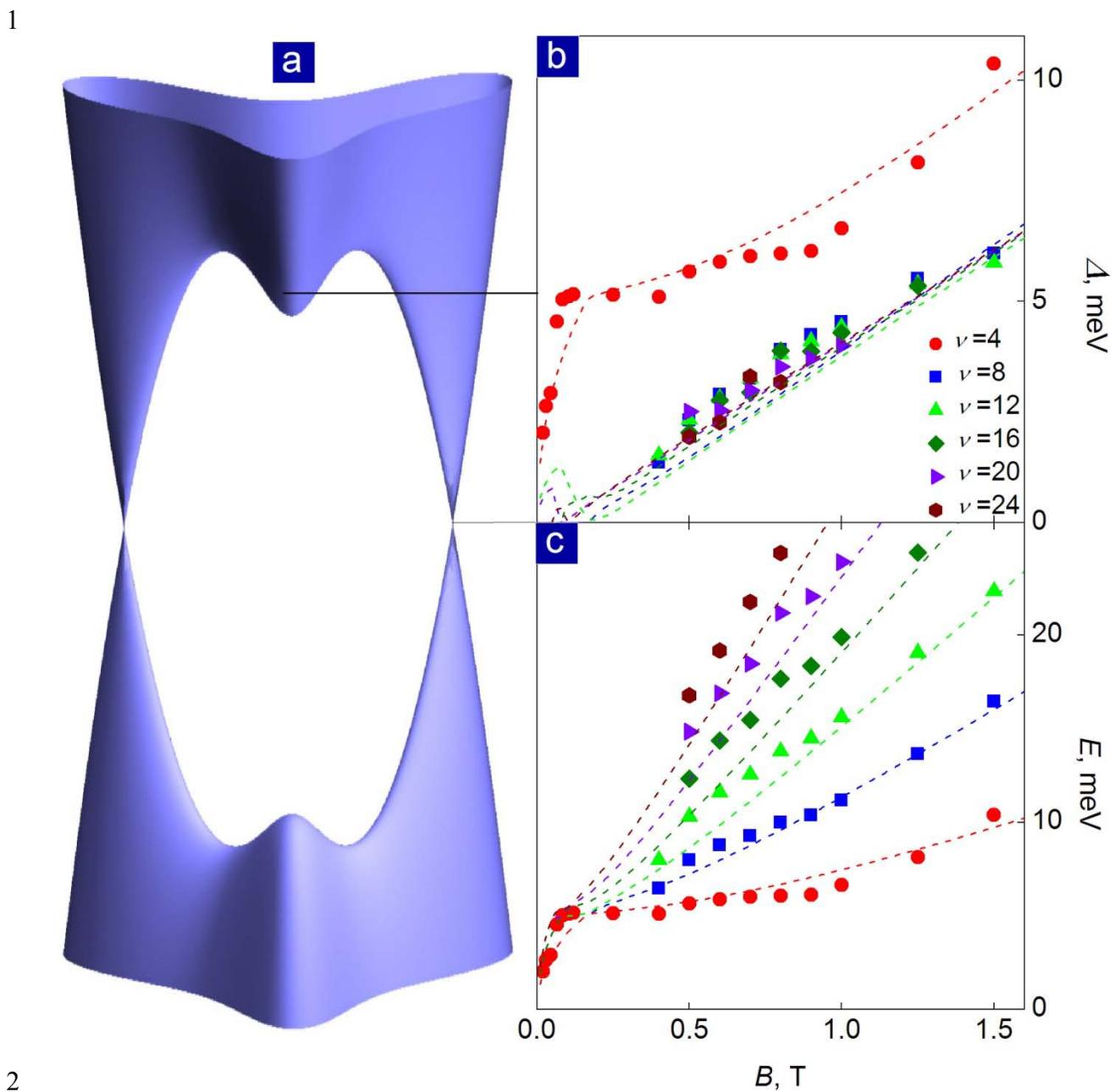

3    Fig. 4

# Interaction-Driven Spectrum Reconstruction in Bilayer Graphene

(Supplementary Information)

A. S. Mayorov[1], D. C. Elias[1], M. Mucha-Kruczynski[2], R. V. Gorbachev[3], T. Tudorovskiy[4], A. Zhukov[3], S. V. Morozov[5], M. I. Katsnelson[4], V. I. Fal'ko[2], A. K. Geim[3], K. S. Novoselov[1]

[1]*School of Physics & Astronomy, University of Manchester, Manchester M13 9PL, UK*

[2]*Physics Department, Lancaster University, Lancaster LA1 4YB, UK*

[3]*Manchester Centre for Mesoscience & Nanotechnology, University of Manchester, Manchester M13 9PL, UK*

[4]*Radboud University of Nijmegen, Institute for Molecules and Materials, Heijendaalseweg 135, 6525 AJ Nijmegen, The Netherlands*

[5]*Institute for Microelectronics Technology, 142432 Chernogolovka, Russia*

# Thermal Broadening of the Conductivity Minimum

Neglecting electron-electron interaction, one can write the static conductivity in the form:

$$\sigma(T,\xi) = \int_{-\infty}^{\infty} K(E)\left(-\frac{\partial f(E,T,\xi)}{\partial E}\right)dE, \tag{1}$$

where $f(E,T,\xi)$ is the Fermi distribution function

$$f(E,T,\xi) = \frac{1}{1+e^{(E-\xi)/T}}, \tag{2}$$

$\xi$ is a chemical potential and $K(E)$ is an unknown function of the energy only to be defined[1]. The expression (1) is valid for any elastic scattering mechanisms, it follows from the general Kubo formula under the assumption that the Hamiltonian and the current density are sums of single-particle operators. It can be rewritten as

$$\sigma(T,\xi) = K(\xi) + \int_0^{\infty} \frac{dx}{1+e^x}\frac{\partial}{\partial x}\left[K(\xi+xT)+K(\xi-xT)\right] \tag{3}$$

The experimentally measured dependence is $\sigma(T) = \sigma(T,\xi(T))$, where the temperature dependence of the chemical potential can be found from the condition, that the concentration *n* persists. The latter can be computed as $n_e - n_h$, where

$$n_e = \int_0^{\infty} N(E)f(E,T,\xi)dE, \quad n_h = \int_{-\infty}^0 N(E)[1-f(E,T,\xi)]dE \tag{4}$$

are total numbers of electrons and holes respectively and $N(E)$ is the density of states. We find

$$n(T,\xi) = \int_0^{\infty} N(E)\left(\frac{1}{1+e^{(E-\xi)/T}} - \frac{1}{1+e^{(E+\xi)/T}}\right)dE, \tag{5}$$

where we assumed that $N(E)$ is even (electron-hole symmetry). Thus

$$n(T,\xi(T)) = \int_0^{\infty} N(E)\left(\frac{1}{1+e^{(E-\xi(T))/T}} - \frac{1}{1+e^{(E+\xi(T))/T}}\right)dE = n = \text{const.} \tag{6}$$

At zero temperature $\sigma(T=0,\xi) = K(\xi)$. On the other hand, *experimentally*, $\sigma(T=0,\xi_0) \propto |n(T=0,\xi_0)|$, where $\xi_0 = \xi(0)$. This gives

$$\sigma(T=0,\xi_0) = K(\xi_0) = \beta |n(T=0,\xi_0)| = \beta \int_0^{|\xi_0|} N(E)dE, \quad (7)$$

where $\beta$ does not depend on $\xi_0$. Equation (7) defines $K(\xi)$. Substituting (7) into (3) we obtain for the conductivity

$$\sigma(T,\xi) = \beta \int_0^{|\xi_0|} N(E)dE$$
$$+ \beta T \int_0^\infty \frac{dx}{1+e^x} [N(\xi+xT)\operatorname{sgn}(\xi+xT) - N(\xi-xT)\operatorname{sgn}(\xi-xT)]. \quad (8)$$

Conductivity (8) can also be rewritten as

$$\sigma(T,\xi) = \beta \int_0^\infty N(E)dE \left( \frac{1}{1+e^{(E-\xi)/T}} + \frac{1}{1+e^{(E+\xi)/T}} \right). \quad (9)$$

The last formula is very natural: the conductivity is proportional to the total number of charge carriers. For a single layer graphene $N(E) = 2|E|/\pi(\hbar V)^2$ and for a bilayer (neglecting trigonal warping effects) $N(E) = 2m/\pi\hbar^2$. From (8) one has

$$\sigma(T,\xi) = \beta_s \left( \xi^2 + \frac{\pi^2}{3}T^2 \right) \quad (10)$$

for the single layer and

$$\sigma(T,\xi) = \beta_b \left( |\xi| + 2T \ln(1+e^{-|\xi|/T}) \right) \quad (11)$$

for the bilayer. For the single layer (6) takes the form

$$\int_0^\infty \left( \frac{1}{1+e^{x-y}} - \frac{1}{1+e^{x+y}} \right) x\,dx = \frac{\xi_0^2}{2T^2}, \quad (12)$$

where $y = \xi(T)/T$. The last equation should be solved numerically. For the bilayer (6) gives

$$n(T,\xi) = \frac{2m|\xi|}{\pi\hbar^2} = \frac{2m|\xi_0|}{\pi\hbar^2}. \quad (13)$$

Thus the chemical potential for the bilayer does not depend on temperature.

## Procedure for extracting of the peak broadening

Let us consider low concentrations $n \to 0$. From (6) we see that in this case $\xi(T)/T \to 0$. Then the conductivity (9) can approximately be written as

$$\sigma(T,\xi) \approx \beta \left[ 2n_T(T) + \xi^2 \left( \frac{N(0)}{4T} + \frac{N'(0)}{2} + \int_0^\infty \frac{N''(E)}{1+e^{E/T}} dE \right) \right], \tag{14}$$

where

$$n_T(T) = \int_0^\infty \frac{1}{1+e^{E/T}} N(E) dE. \tag{15}$$

We note that the relation (14) holds for any density of states $N(E)$. This allows us to reconstruct of $N(E)$ from conductivity measurements. For the single layer

$$n_T(T) = \frac{\pi T^2}{6(\hbar V)^2}. \tag{16}$$

For the bilayer

$$n_T(T) = \frac{2mT}{\pi \hbar^2} \ln 2. \tag{17}$$

For low concentrations (6) gives

$$n \approx \xi(T) \left( N(0) + 2\int_0^\infty \frac{N'(E)}{1+e^{E/T}} dE \right). \tag{18}$$

This gives

$$\xi(T) = \frac{\pi(\hbar V)^2}{4T \ln 2} n, \quad \xi(T) = \frac{\pi \hbar^2}{2m} n. \tag{19}$$

for single and bilayer respectively. Finally for the single layer

$$\sigma(T) \approx 2\beta n_T(T) \left[ 1 + \frac{\pi^2}{192(\ln 2)^2} \frac{n^2}{n_T^2(T)} \right], \tag{20}$$

and for the bilayer

$$\sigma(T) \approx 2\beta n_T(T) \left[ 1 + \frac{\ln 2}{8} \frac{n^2}{n_T^2(T)} \right]. \tag{21}$$

For the resistance we have $R(T) = \sigma^{-1}(T)$. One sees that for low concentrations resistance takes a form of a Lorentzian with a width

$$\Gamma(T) = \frac{8\ln 2}{\sqrt{3}(\hbar V)^2} T^2 \qquad (22)$$

for the single layer and

$$\Gamma(T) = \frac{8m\sqrt{2\ln 2}}{\pi\hbar^2} T. \qquad (23)$$

for the bilayer. At zero concentration the resistance decays as $T^{-2}$ for a single layer and as $T^{-1}$ for a bilayer.

At zero concentration we fit our experimental data by

$$\sigma(T) = \alpha + 2\beta n_T(T) \qquad (24)$$

to find $n_T(T)$. Here $\alpha$ takes into account the minimum conductivity at zero temperature. The dependence $n_T(T)$ is plotted in Fig. 2. For a bilayer we expect $n_T(T)$ to obey (17), from what the effective mass can be extracted (see the main text).

## Energy spectrum in magnetic field

The Hamiltonian, describing the bi-layer graphene quasiparticle dynamics reads [2]:

$$H = \begin{pmatrix} 0 & \dfrac{\pi_-^2}{2m} + v_3\pi_+ \\ \dfrac{\pi_+^2}{2m} + v_3\pi_- & 0 \end{pmatrix}, \qquad (25)$$

where $\pi_\pm = \pi_x \pm i\pi_y$, $\pi_j = -i\hbar\nabla_j + (e/c)A_j$, $j = x, y$, $e > 0$ is the absolute value of electron charge, $c$ is the speed of light, $A$ is the vector-potential. We introduce creation-annihilation Bose-operators $b^+$, $b$ as follows

$$\pi_+ = \sqrt{2e\hbar B/c}\, b^+, \quad \pi_- = \sqrt{2e\hbar B/c}\, b. \qquad (26)$$

They satisfy the following commutation relation

$$[b, b^+] = 1. \qquad (27)$$

The spectrum can be found from the equation $H\psi = E\psi$, $\psi = (\psi_1, \psi_2)$. Excluding $\psi_1$ from the last equation we obtain

$$[(b^+)^2 b^2 + \alpha((b^+)^3 + b^3) + \alpha^2 bb^+]\psi_2 = \varepsilon^2\psi_2, \qquad (28)$$

where $\alpha = mv_3\sqrt{2c/(e\hbar B)}$, $\varepsilon = E/(\hbar\omega_B)$, $\omega_B = eB/(mc)$. In the basis of eigenfunctions $|n\rangle$ of the operator $b^+ b$ this equation takes the form $M\eta = \varepsilon^2\eta$, where $\eta = (\eta_0, \eta_1, ...)$ and the matrix $M$ is

$$M_{nn'} = [n(n-1) + \alpha^2(n+1)]\delta_{nn'} + \tag{29}$$
$$\alpha\left[\sqrt{(n+1)(n+2)(n+3)}\delta_{n+3,n'} + \sqrt{n(n-1)(n-2)}\delta_{n-3,n'}\right]$$

Eigenvalues $\varepsilon^2$ of the sparse matrix $M$ can be found using the Arnoldi method with a shift [3]. The shift is required due to the degeneracy of the zero eigenvalue.

For a weak magnetic field $\alpha \ll 1$ and the distortion of Landau levels can be found analytically. Indeed we write $(H_0 + H_1)\varphi = \varepsilon^2 \varphi$, where $H_0 = (b^+)^2 b^2 + \alpha^2 b b^+$ and $H_1 = \alpha((b^+)^3 + b^3)$. The perturbation theory gives $z_{n,0} = n(n-1) + \alpha^2(n+1)$, $z_{n,1} = 0$,

$$z_{n,2} = \sum_{n \neq m} \frac{|\langle n | H_1 | m \rangle|^2}{z_{n,0} - z_{m,0}}, \tag{30}$$

Using (24) we find $z_{n,2} = -\alpha^2(n+1)$, $z_{n,0} + z_{n,2} = n(n-1)$, i.e. eigenvalues are not perturbed up to $1/B$. The previous computation is valid for $n > 2$. For $n \leq 2$

$$z_{n,0} + z_{n,2} = n(n-1) + \alpha^2 n(1-n)/6. \tag{31}$$

Thus for $n = 0,1$ we still get the absence of $1/B$ correction. The only case when the correction arises is $n = 2$. Then $z_{n,0} + z_{n,2} = 2 - \alpha^2/3$, $E \approx \pm(\sqrt{2}\hbar\omega_B - \Delta E)$. The estimation gives $\Delta E = -\sqrt{2}mv_3^2/6 = -0.81\,\text{meV}$. Thus

$$E_n \approx \pm\left(\hbar\omega_B\sqrt{n(n-1)} - \Delta E \delta_{n,2}\right). \tag{32}$$

Level spacings $E_{n+1} - E_n$ corresponding to the plus sign for $n = 1,...,4$ are $1.414\hbar\omega_B - \Delta E$, $1.035\hbar\omega_B + \Delta E$, $1.015\hbar\omega_B$. Level spacings for higher levels rapidly tend to $\hbar\omega_B$ since $E_n \approx \hbar\omega_B$. Thus, all level spacings are approximately equal to $\hbar\omega_B$ apart from $n = 0,1,2$.

## Thermodynamic density of states and effective mass

Oscillations of magnetoresistance (Shubnikov-de Haas) and magnetization (Haas-van Alphen) caused by magnetic field are closely related to the "thermodynamic density of states" (TDOS)

$$D(\xi) = \int_{-\infty}^{\infty} N(E) \frac{\partial f(E-\xi)}{\partial \xi} dE, \tag{33}$$

where $\xi$ is the chemical potential counted from the neutrality point. The Landau levels (LL) for bilayer graphene are $\pm E_n$, $E_n = \varepsilon_c \sqrt{n(n+1)}$, $\varepsilon_c = \hbar\omega_B$. Taking into account a broadening of LL, we have

$$n(E) = g_s g_v \frac{\Phi}{\Phi_0}\left\{2\frac{\Gamma}{\pi}\frac{1}{E^2+\Gamma_0^2} + \sum_{n=1}^{\infty}\left[\frac{\Gamma}{\pi}\frac{1}{(E-E_n)^2+\Gamma^2} + \frac{\Gamma}{\pi}\frac{1}{(E+E_n)^2+\Gamma^2}\right]\right\}, \tag{34}$$

where $g_s = g_v = 2$ are spin and valley degeneracies, $\Phi, \Phi_0 = \hbar c/e$ are the total flux per sample and flux quantum, respectively and $\Gamma$ is the broadening of LL. Factor $2$, corresponding to $E_0 = 0$ reflects the degeneracy of the zero level. The Fourier component of $\partial f(E-\xi)/\partial \xi$ can be computed using the residue theory:

$$\int_{-\infty}^{\infty} dE e^{iEt} \frac{\partial f(E-\xi)}{\partial \xi} = e^{it\xi} \frac{\pi T t}{\sinh(\pi T t)}. \tag{35}$$

Then

$$D(\xi) = \int\int_{-\infty}^{\infty} \frac{dEdt}{2\pi} e^{i(\xi-E)t} \frac{\pi T t}{\sinh(\pi T t)} N(E). \tag{36}$$

We note that $N(E)$ can be written in the form

$$N(E) = g_s g_v \frac{\Phi}{\Phi_0} \frac{\Gamma}{2\pi} \sum_{n=-\infty}^{\infty} \left[ \frac{1}{(E-E_n)^2 + \Gamma^2} + \frac{1}{(E+E_n)^2 + \Gamma^2} \right]. \tag{37}$$

From the Poisson summation rule

$$\sum_{n=-\infty}^{\infty} f(n) = \sum_{k=-\infty}^{\infty} \int_{-\infty}^{\infty} e^{2\pi i k x} f(x) dx \tag{38}$$

we obtain

$$N(E) = g_s g_v \frac{\Phi}{\Phi_0} \frac{\Gamma}{2\pi} \sum_{k=-\infty}^{\infty} \int_{-\infty}^{\infty} e^{2\pi i k x} f(x) dx, \tag{39}$$

where

$$f(n) = \frac{\Gamma}{\pi} \frac{E_n^2 + E^2 + \Gamma^2}{(E_n^2 - z_+^2)(E_n^2 - z_-^2)}, \tag{40}$$

and $z_\pm = E \pm i\Gamma$. The residue theory gives

$$\int_{-\infty}^{\infty} e^{2\pi i k x} f(x) dx = \frac{(-1)^k}{\varepsilon_c^2} \left[ \frac{z_+}{\alpha(z_+)} e^{i\pi |k| \alpha(z_+)} + \frac{z_-}{\alpha(z_-)} e^{i\pi |k| \alpha(z_-)} \right], \tag{41}$$

where $\alpha(z) = \sqrt{4z^2/\varepsilon_c^2 + 1}$. The oscillating part of TDOS is

$$D_{osc}(\mu) = \frac{g_s g_v}{\pi \varepsilon_c^2} \frac{\Phi}{\Phi_0} \int\int_{-\infty}^{\infty} e^{i(\xi-E)t} \frac{\pi T t}{\sinh(\pi T t)} \sum_{k=1}^{\infty} (-1)^k [R_+(k,E) + R_-(k,E)] dt dE$$
$$, \tag{42}$$

$$R_\pm(k,E) = \frac{z_\pm}{\alpha(z_\pm)} e^{\pm i\pi k \alpha(z_\pm)}.$$

Assuming that $S(t,E) = (\xi - E)t + \pi k \alpha(E) >> 1$, $\Gamma << \xi$ we can use the stationary phase method to compute $D_{osc}(\xi)$ [4]. The result is

$$D_{osc}(\xi) = \frac{2g_s g_v}{\varepsilon_c^2} \frac{\Phi}{\Phi_0} \frac{1}{\sqrt{1+\varepsilon_c^2/4\xi^2}} \sum_{k=1}^{\infty} (-1)^k Z(k) \cos\left(\frac{2\pi k \mu}{\varepsilon_c}\sqrt{1+\frac{\varepsilon_c^2}{4\xi^2}}\right) e^{-\pi k \beta \Gamma}, \quad (43)$$

$$Z(k) = \frac{\pi kT\beta}{\sinh(\pi kT\beta)}, \quad \beta = \frac{2\pi m^* c}{\hbar eB}, \quad m^* = \frac{m}{\sqrt{1+\varepsilon_c^2/4\xi^2}}.$$

## Minimum conductivity of graphene and bilayer graphene

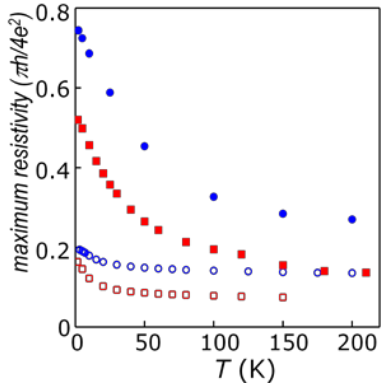

Fig. S1. Resistivity at the compensation point for two monolayer samples (solid symbols) and two bilayer devices (open symbols).

Our suspended samples, being nearly ballistic, demonstrate very different temperature dependence of the conductivity minimum in comparison to its non-suspended counterparts [7-8]. The minimum resistivity in low-mobility graphene on the substrate is usually higher or close to $4e^2/h$ [7-8], which is a factor of $\pi$ larger than that predicted by theory for ballistic transport [9-10]. Interestingly, the minimum conductivity for our suspended samples is approaching its theoretical value of $4e^2/\pi h$ (Fig. S1) which is in agreement with the previous results [11].

For ballistic bilayer graphene with parabolic dispersion relation the predicted minimum conductivity is $8e^2/\pi h$ - twice that of monolayer graphene [12-13], whenever, if the effect of the trigonal warping is taken into account, the minimum conductivity is expected to be six times that of monolayer - $24e^2/\pi h$ (*4*). Our experimental results demonstrate some intermediate value of about $20e^2/\pi h$, which might indicate the reduced number of pockets with linear dispersion around each of the K(K') points.

## Another Example of Bilayer Graphene Device

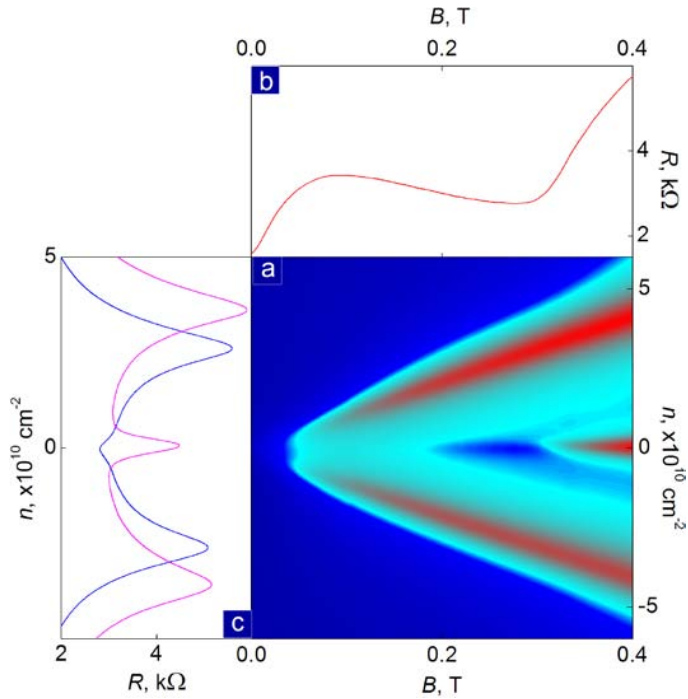

Fig. S2. Free-standing bilayer samples in magnetic field. a, Contour map of resistance of free-standing bilayer graphene device as a function of carrier concentration and magnetic field. The resistance spans from 2.1kΩ (blue) to 6kΩ (red). b, Resistance of suspended bilayer graphene at $\nu=0$ as a function of magnetic field. Note negative differential magnetoresistance at approx 0.1T<$B$<0.3T. c, Resistance of suspended bilayer graphene as a function of gate voltage (recalculated to carrier concentration) for B=0.25T (blue); and $B$=0.35T (magenta). Note the sharp minimum on $B$=0.25T curve at -1×10$^9$ cm$^{-2}$<$n$<1×10$^9$ cm$^{-2}$, which is replaced by a local maximum on $B$=0.35T curve.

To demonstrate how ubiquitous is the behavior we observed in our bilayer samples we present data from another bilayer sample with slightly lower mobility (~700,000cm$^2$/V·s). All the characteristic features we observed for the sample presented in the main text can be seen here as well: (i) the resistance at filling factor zero is a nonmonotonous function of the magnetic field and diverges above 0.4T, which indicates the presence of a gap in the spectrum; (ii) the cyclotron gaps at filling factors ±4 (maxima in the two-probe resistance) is significantly larger than the cyclotron gaps at other filling factors (so the oscillation at $\nu=\pm4$ survive for very low magnetic fields). The extracted gaps for different filling factors follow similar pattern presented in the main text Fig. 4, just the plateau feature observed for $\nu=\pm4$ is at slightly lower energy (~3meV).

## Resonance-like dip at filling factor zero

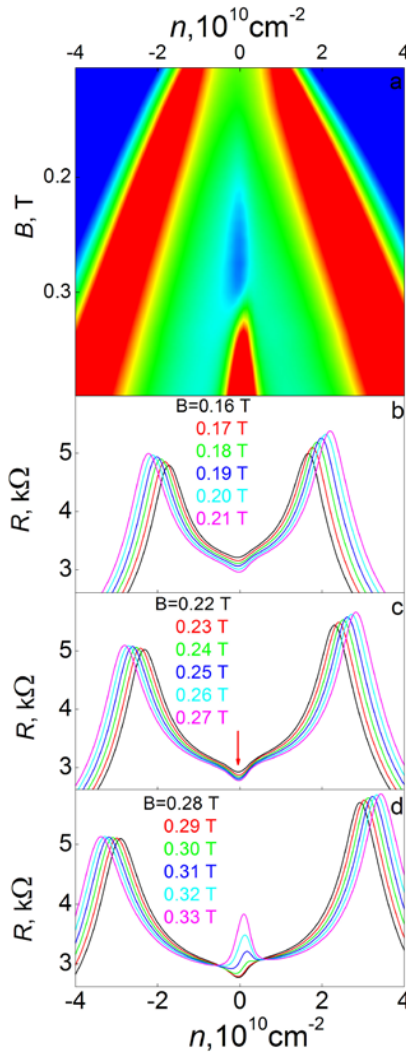

Fig. S3. A sharp local minimum (marked by the red arrow) appears for carrier concentrations $3\times10^9$ cm$^{-2}$ < n < $3\times10^9$ cm$^{-2}$ and magnetic fields 0.18T<B<0.3T.

Magnetoresistance in our samples is a strongly non-monotonous function of magnetic field. For instance, for the samples presented on Fig. S2, magnetoresistance is positive for B<0.1T, negative for 0.1T<B<0.3T and positive for B>0.3T. Moreover, in the area of negative magentoresistance a sharp, resonance-like dip is observed for carrier concentration -$3\times10^9$ cm$^{-2}$ < n < $3\times10^9$ cm$^{-2}$ (Fig. S2, S3).

The negative magnetoresistance can be explained by the increase of the density of states in magnetic field at the compensation point due to formation of zero Landau level (at B=0 the density of states at the Dirac point is zero for the linear spectrum). Thus, the conductivity should rise from the quantum limited value (for the unreconstructed spectrum it was estimated [5-6] as $24e^2/\pi h$) to that determined by the density of states at the zero Landau level.

At the moment we haven't got an explanation for the appearance of the sharp dip in the resistance, but would like to point out that the range of carrier concentration occupied by the dip corresponds roughly to the number of states for within the conical part of the bilayer spectrum.